\documentstyle[epsf]{mn}

\newif\ifAMStwofonts
\AMStwofontstrue

\makeatletter
\def\figsize{\ifSFB@referee\epsfxsize=0.5\hsize\else\epsfxsize=\hsize\fi}
\makeatother
\gdef\0{\phantom{0}}
\def\n(#1){{}$^{\rm(#1)}$}
\def\eq#1 {\begin{equation} #1 \end{equation}}
\def\eqarray#1{\begin{eqnarray} #1 \end{eqnarray}}
\def\Ref{\bibitem{}}
\def\symbol#1{\hbox{$#1$}}
\def\about{\symbol{\sim}}

\def\nhat {\symbol{\hat n}}
\def\Mo   {{\hbox{M$_\odot$}}}
\def\Lo   {{\hbox{L$_\odot$}}}
\def\Tstar{\symbol{T_\ast}}
\def\Mdot {{\hbox{$\dot M$}}}
\def\kms  {{\hbox{km s$^{-1}$}}}
\def\mic  {{\hbox{$\umu$m}}}
\ifoldfss
  \ifCUPmtlplainloaded \else
    \NewTextAlphabet{textbfit} {cmbxti10} {}
    \NewTextAlphabet{textbfss} {cmssbx10} {}
    \NewMathAlphabet{mathbfit} {cmbxti10} {} % for math mode
    \NewMathAlphabet{mathbfss} {cmssbx10} {} %  "   "    "
  \fi
  \ifAMStwofonts
    \ifCUPmtlplainloaded \else
      \NewSymbolFont{upmath} {eurm10}
      \NewSymbolFont{AMSa} {msam10}
      \NewMathSymbol{\upi}     {0}{upmath}{19}
      \NewMathSymbol{\umu}     {0}{upmath}{16}
      \NewMathSymbol{\upartial}{0}{upmath}{40}
      \NewMathSymbol{\leqslant}{3}{AMSa}{36}
      \NewMathSymbol{\geqslant}{3}{AMSa}{3E}

       \let\le=\leqslant
       \let\ge=\geqslant
    \fi
  \fi
\fi % End of OFSS

\ifnfssone
  \newmathalphabet{\mathit}
  \addtoversion{normal}{\mathit}{cmr}{m}{it}
  \addtoversion{bold}{\mathit}{cmr}{bx}{it}
  \newmathalphabet{\mathbfit} % math mode version of \textbfit{..}
  \addtoversion{normal}{\mathbfit}{cmr}{bx}{it}
  \addtoversion{bold}{\mathbfit}{cmr}{bx}{it}
  \newmathalphabet{\mathbfss} % math mode version of \textbfss{..}
  \addtoversion{normal}{\mathbfss}{cmss}{bx}{n}
  \addtoversion{bold}{\mathbfss}{cmss}{bx}{n}
  \ifAMStwofonts
    \ifCUPmtlplainloaded \else
      \UseAMStwoboldmath
      \makeatletter
      \new@mathgroup\upmath@group
      \define@mathgroup\mv@normal\upmath@group{eur}{m}{n}
      \define@mathgroup\mv@bold\upmath@group{eur}{b}{n}
      \edef\UPM{\hexnumber\upmath@group}
      \new@mathgroup\amsa@group
      \define@mathgroup\mv@normal\amsa@group{msa}{m}{n}
      \define@mathgroup\mv@bold\amsa@group{msa}{m}{n}
      \edef\AMSa{\hexnumber\amsa@group}
      \makeatother
      \mathchardef\upi="0\UPM19
      \mathchardef\umu="0\UPM16
      \mathchardef\upartial="0\UPM40
      \mathchardef\leqslant="3\AMSa36
      \mathchardef\geqslant="3\AMSa3E

       \let\le=\leqslant
       \let\ge=\geqslant
    \fi
  \fi
\fi % End of NFSS release 1

\ifnfsstwo
  \DeclareMathAlphabet{\mathbfit}{OT1}{cmr}{bx}{it}
  \SetMathAlphabet\mathbfit{bold}{OT1}{cmr}{bx}{it}
  \DeclareMathAlphabet{\mathbfss}{OT1}{cmss}{bx}{n}
  \SetMathAlphabet\mathbfss{bold}{OT1}{cmss}{bx}{n}
  \ifAMStwofonts
    \ifCUPmtlplainloaded \else
      \DeclareSymbolFont{UPM}{U}{eur}{m}{n}
      \SetSymbolFont{UPM}{bold}{U}{eur}{b}{n}
      \DeclareSymbolFont{AMSa}{U}{msa}{m}{n}
      \DeclareMathSymbol{\upi}{0}{UPM}{"19}
      \DeclareMathSymbol{\umu}{0}{UPM}{"16}
      \DeclareMathSymbol{\upartial}{0}{UPM}{"40}
      \DeclareMathSymbol{\leqslant}{3}{AMSa}{"36}
      \DeclareMathSymbol{\geqslant}{3}{AMSa}{"3E}

       \let\le=\leqslant
       \let\ge=\geqslant
    \fi
  \fi
\fi % End of NFSS release 2

\ifCUPmtlplainloaded \else
  \ifAMStwofonts \else % If no AMS fonts
    \def\upi{\pi}
    \def\umu{\mu}
    \def\upartial{\partial}
  \fi
\fi

\pagerange{\pageref{firstpage}--\pageref{lastpage}}
\pubyear{1996}

\title[Dust emission from IRC+10216]
                         {Dust emission from IRC+10216}

\author[\v{Z}. Ivezi\'{c} and M. Elitzur]
                  {\v{Z}eljko Ivezi\'{c} and Moshe Elitzur \\
                        Department of Physics and Astronomy,
        University of Kentucky, Lexington, KY 40506-0055, USA\\
                   e-mail: ivezic@pa.uky.edu, moshe@pa.uky.edu }

\date{Accepted 1995 November 24. Received 1995 November 23;
      in original form 1995 July 31}

\begin{document}

\maketitle

\label{firstpage}

\begin                             {abstract}
Infrared emission from the dust shell around IRC+10216 is analysed in detail,
employing a self-consistent model for radiatively driven winds around late-type
stars that couples the equations of motion and radiative transfer in the dust.
The resulting model provides agreement with the wealth of available data,
including the spectral energy distribution in the range 0.5--1000 \mic, and
visibility and array observations.  Previous conclusions about two dust shells,
derived from modelling the data with a few single-temperature components of
different radii, are not supported by our results. The extended, continuous
temperature and density distributions derived from our model obviate the need
for such discrete shells. The IR properties vary with the stellar phase,
reflecting changes in both the dust condensation radius $r_1$ and the overall
optical depth $\tau$ -- as the luminosity increases from minimum to maximum,
$r_1$ increases while $\tau$ decreases. We find that the angular size of the
dust condensation zone varies from 0.3 arcsec at minimum light to 0.5 arcsec at
maximum.  The shortage of flux at short wavelengths encountered in previous
studies is resolved by employing a grain size distribution that includes grains
larger than \about\ 0.1 \mic, required also for the visibility fits.  This
distribution is in agreement with the one recently proposed by Jura in a study
that probed the outer regions of the envelope. Since our constraints on the
size distribution mostly reflect the envelope's inner regions, the agreement of
these independent studies is evidence against significant changes in grain
sizes through effects like sputtering or grain growth after the initial
formation at the dust condensation zone.
\end{abstract}


\section                    { INTRODUCTION }

IRC+10216 (CW Leo, IRAS 09452+1330) is by far the brightest and best-studied
mass-losing carbon star (Jura \& Kleinmann 1989). Starting with Mitchell \&
Robinson (1980), several authors have performed radiative transfer
calculations for the IR dust emission from this source.  However, with the
exception of Winters, Dominik \& Sedlmayr (1994), all the previous studies
were based on a prescribed $r^{-2}$ radial density distribution that is not
fully consistent with those of radiatively driven winds.  Furthermore,
although reasonable fits to the observed spectral energy distribution (SED)
were generated over a wide range of wavelengths, none of the models produced
enough flux shortward of 1--2 \mic\ (e.g. Le Bertre 1987; Keady, Hall \&
Ridgway 1988; Griffin 1990; Lorenz-Martins \& Lef\' evre 1993).  In addition,
none of the models provides simultaneous agreement with spatially resolved
observations at 2.2 \mic\ (e.g. Martin \& Rogers 1987).

The purpose of this work is to perform a self-consistent study that employs a
dust density distribution determined from the solution of the coupled system
of radiative transfer and hydrodynamics equations for the wind.  The equations
are described elsewhere (Netzer \& Elitzur 1993; Ivezi\'c \& Elitzur 1995,
hereafter IE95). As shown in IE95, the solution of this system is essentially
determined by a single quantity -- the flux-averaged optical depth $\tau_{\rm
F}$. Once $\tau_{\rm F}$ is determined, scaling relations listed in IE95 and in
Ivezi\' c \& Elitzur (1996; hereafter IE96) can be used to constrain all other
relevant quantities. In principle, the optical depth can be estimated from
either the spectral shape $f_\lambda = F_\lambda/F$, where $F = \int F_\lambda
\rm d\lambda$ is the bolometric flux, or spatially resolved observations.
However, since the latter depend also on the angular scale of the system,
because of observational uncertainties the determination of optical depth from
the spectral shape is much more reliable. In IE96 we describe a two-step
modelling procedure, which we follow in this work.  In the first step the dust
characteristics and overall optical depth are constrained from the best fit to
the spectral shape. Then, with the model prediction for the surface brightness
distribution based on these parameters, the spatially resolved observations
are used in the second step to determine the angular size of the dust
condensation zone.

The spectral energy distribution is discussed in Section 2, high-resolution
observations in Section 3 and outflow dynamics in Section 4. The results are
summarized and discussed in Section 5.

\section                { SPECTRAL ENERGY DISTRIBUTION }

IRC+10216 is a long-period variable with a period of 638 d and a recent
minimum at JD = 244\,7863 (Dyck et al.\ 1991).  We limit our analysis to the
periodical changes and do not consider the long-term modulations with
time-scales of a few decades noted by Dyck et al. The surface brightness
distribution at short wavelengths ($\le$ 2--3 \mic) is slightly asymmetric
(e.g. Ridgway \& Keady 1988; Kastner \& Weintraub 1994). However, these
spatial asymmetries decrease as the wavelength increases (e.g.\ de Batz
1988)\footnote{Recently, Sloan \& Egan (1995) observed IRC+10216 with a
$0.9\times2.0$ arcsec$^2$ resolution at 10 \mic\ and obtained an indication of
a blue emission patch with a size of \about\ 1 arcsec, located \about\ 1
arcsec north of the star.  Note that these spatial details are comparable to
the slit size; indeed, Sloan \& Egan point out that this component's location
could be an artefact of the reconstruction algorithm and that it could
actually originate from the inner region centered on the star.  Such emission
indeed is expected from this region due to the hot dust. These observations do
not seem to refute the findings of de Batz.} and our model assumes spherical
symmetry. Slight disagreement between the model and observations can be
expected at short wavelengths, a point further discussed in Section 5.

%%%%%%%%%%%%%%%%%%% Figure 1 %%%%%%%%%%%%%%%%%%%%%%%%%%%%%
\begin{figure}
%\centering \leavevmode \figsize \epsfbox[55 65 535 765]{ircsed.ps}

\caption{Spectral energy distribution for IRC +10216; lines represent model
results, symbols the observations.  Data are from Le Bertre (1987) (\vrule
height 3pt width 3pt depth 0pt), (1988) ($\circ$); Rengarajan et al.\ (1985)
($\bullet$); and {\it IRAS} Point Source Catalogue ($\star$).  All observations
are at maximum light except for those denoted by open circles, which were at
minimum light. The thick solid line is the model result for maximum light, the
thin solid line the result for minimum; details are described in the text. The
dashed line is the model result for maximum light and single-size (0.05 \mic)
grains. The inset shows an expanded view of the {\it IRAS} LRS spectral region
-- the dots are the data, taken close to maximum light, the solid line the
model.}

\end{figure}
%%%%%%%%%%%%%%%%%%%%%%%%%%%%%%%%%%%%%%%%%%%%%%%%%%%%%%%%%%%%%%%%%%%%%%%%%%%%

Our best-fitting model to the spectral shape is shown in Fig. 1 together with
the observations.  The thick solid line corresponds to maximum light, the thin
solid line to minimum (where there are only five observational points). The
inset compares the model results (solid line) with the {\it IRAS} LRS data.
The model is primarily determined by the overall optical depth and the dust
composition.\footnote{The required grain properties are the spectral shapes of
the absorption and scattering efficiencies. Absolute values of these
quantities are not needed.} From previous work (e.g. Blanco et al.\ 1994), the
dust grains around IRC+10216 are primarily composed of amorphous carbon with a
minor inclusion of SiC to account for the 11.3-\mic\ feature. With optical
properties for amorphous carbon taken from Hanner (1988) and for SiC from P\'
egouri\' e (1988), we find that the best fit to the 11.3-\mic\ feature is
obtained with a mixture of 95 per cent amorphous carbon and 5 per cent SiC (by
mass), although varying the percentage of SiC in the range 3--8 per cent still
produces satisfactory agreement.  Griffin (1990) presents results for various
SiC abundances. He obtains the best agreement for 17 per cent SiC, but the
quality of his fit at 8 per cent is comparable.  The slight difference between
the conclusions of the two studies is insignificant and might be explained by
the $r^{-2}$ density law employed by Griffin instead of the hydrodynamic
calculation done here.  A broad emission feature between 24 and 30 \mic\
(Forrest, Houck \& McCarthy 1981) provides evidence for an additional
component, probably MgS compound (Goebel \& Moseley 1980).  With the aid of
spectroscopic data for MgS from Nuth et al.\ (1985) we estimate the abundance
of this component to be less than 10 per cent, if this chemical identification
is correct.

In addition to the chemical composition, the distribution of grain radii $a$
also affects the optical properties.  However, the wavelength dependence of
absorption and scattering efficiencies is independent of $a$ once $\lambda \ga
2\upi a$. Therefore, at the wavelengths of interest, $\lambda \ge 0.5$ \mic,
the grain size is irrelevant as long as $a \la 0.1$ \mic, and models of
IRC+10216 usually assumed that all grains have the same size $a=0.05$ \mic\
(for an overview see Lorenz-Martins \& Lef\' evre 1993). The dashed line in
Fig.\ 1 shows our model result for this single size, displaying the problems
encountered by all other workers -- the models do not produce enough flux at
$\lambda \le$ 1 \mic.

What is the meaning of this discrepancy? Dust emission is insignificant at
$\lambda \la$ 3 \mic\ because it decreases exponentially for wavelengths
shorter than $3~\mic\times(1000\,{\rm K}/T_1)$, the peak wavelength of the
Planck distribution for the dust condensation temperature $T_1$, the highest
possible dust temperature. Therefore, the detected radiation involves only
attenuated stellar emission and scattered light.  Our detailed models show
that, under these conditions, the spectral shape is proportional to
exp$(-\tau_{\rm abs})$, where $\tau_{\rm abs}$ is the overall optical depth
for absorption. Therefore, the shortage of observed flux at $\lambda \la$ 1
\mic\ implies that the model estimates for $\tau_{\rm abs}(\lambda)$ at these
wavelengths are too large.  Since in general $\tau_{\rm abs}(\lambda)$
increases as the wavelength decreases, this rise must be suppressed around 1
\mic. Indeed, Rowan-Robinson \& Harris (1983) noted that by postulating a
departure from the usual $\lambda^{-1}$ dependence of $\tau_{\rm
abs}(\lambda)$ to a flatter distribution at $\lambda \le$ 1 \mic\ they could
produce a better agreement with observations. Although they did not attempt to
justify this behaviour, it can be modelled by assuming a range of grain sizes
$a$. While $Q_{\rm abs} \propto \lambda^{-1}$ for $\lambda \ga 2\upi a$, it is
approximately constant for $\lambda <2 \upi a$. Consequently, the flux
shortage at $\lambda \la$ 1 \mic\ can be alleviated by adding larger grains
with sizes of $a \simeq \lambda/2\upi \approx$ 0.2 \mic, suppressing the rise
of $\tau_{\rm abs}(\lambda)$ when $\lambda$ decreases below \about\ 1 \mic.

For the detailed models we employed two types of size distributions $n(a)$.
Most often used is
\eq{\label{MRN}
             n_{\rm MRN}(a) \propto a^{-3.5}, \qquad a \le a_{\rm max}
}
proposed by Mathis, Rumpl \& Nordsieck (1977).  This MRN distribution includes
a sharp cutoff, $a_{\rm max}$, to the grain radii, required by the finite amount
of mass in the dust. Recently Jura (1994) proposed a modification of the form
\eq{\label{Jura}
                n_{\rm J}(a) \propto a^{-3.5} {\rm e}^{-a/a_0},
}
replacing the sharp cutoff with an exponential one.  Both distributions can
produce satisfactory fits: the MRN distribution requires $a_{\rm max} \approx$
0.2--0.3 \mic, the Jura distribution $a_0 \approx$ 0.15--0.2 \mic. It appears
that the Jura distribution produces a slightly better fit to the spectral
shape observed at maximum light, but any stronger conclusion is hampered by
the observational uncertainties.  In contrast, recent models by Bagnulo, Doyle
\& Griffin (1995) produced satisfactory fits with single-size grains of 0.02
\mic\ as well as the MRN distribution with $a_{\rm max} = 0.05$ \mic, but not
with the Jura distribution.  However, these models used an $r^{-2}$ density
law instead of employing a self-consistent hydrodynamic calculation, as done
here. The two density distributions are substantially different at the inner
regions -- the self-consistent distribution has a much faster initial fall-off
and has already dropped by a factor of 5 below the $r^{-2}$ distribution at $r
\sim 1.5r_1$. Since this is the region where the short wavelengths are
produced, this could account for the different results.

We have thus determined the two major ingredients that affect the spectral
shape, the grain optical properties and overall optical depth.  In addition,
the stellar temperature $T_\ast$ and dust condensation temperature $T_1$ have
a discernible effect on the spectral shape, but only at short wavelengths. Our
best fit gives $T_\ast = 2200 \pm 150$ K, an estimate in agreement with a
spectral type of C9 (Cohen 1979) and the majority of other models.  In
general, the effect of \Tstar\ is limited to $\lambda \la$ 4 \mic\ and its
significance is diminished as the envelope's optical depth increases. Our
best-fitting estimate for $T_1$ is $750 \pm 50$ K. The effect of $T_1$ is more
significant because this parameter controls the peak wavelength of the
spectral shape in envelopes that are optically thin around that peak. Our
estimate for $T_1$, determined from the observed spectral shape by the
location of the peak and the sharp decline toward short wavelengths, is
somewhat lower than the \about\ 1000 K obtained in most other models. Indeed,
in these models the peak of the spectral shape is shifted slightly to the
left, resulting in excessive flux in the 2--7 \mic\ wavelength range (e.g., Le
Bertre 1987; Lorenz-Martins \& Lef\' evre 1993).

Finally, the radius of the envelope's outer edge, $r_{\rm out}$, must be
specified for a numerical solution. Because of scaling, only the relative
thickness $y_{\rm out} = r_{\rm out} / r_1$ is needed.  This parameter affects
only the long-wavelength part of the solution, which is afflicted by a number
of uncertainties.  First, the behaviour of the absorption efficiency is quite
uncertain at these wavelengths. Usually modelled by a power law $Q_{\rm abs}
\propto \lambda^{-\beta}$, the value of $\beta$ is poorly known, typically
taken as \about\ 1--1.5. Next, the long-wavelength tail of the SED could
contain a significant contribution from free--free emission (Griffin 1990).
Fortunately, apart from the long wavelength part of the SED, the model results
are not very sensitive to these uncertainties.  We find from fits to the
spectral shape in the wavelength range 100--1000 \mic\ that $\beta$ varies
from 1.2 to 1.6 for $y_{\rm out}$ between 600 and 10\,000. An independent
estimate for $y_{\rm out}$ can be obtained from the extent of molecular
emission. CO observations by Huggins, Olofsson \& Johansson (1988) indicate
that $y_{\rm out} \ge 700$, and consequently $\beta \ge 1.2$. Indeed, Jura
(1983) suggested that $\beta = 1.3$, a proposal supported by Le Bertre (1987).
The results presented in Fig. 1 are for $y_{\rm out} =700$ and $\beta =1.3$.
For this $\beta$, $y_{\rm out}$ can be increased all the way to 10\,000
without a significant degradation of the fits.

%%%%%%%%%%%%%%%%%%%% Table 1 %%%%%%%%%%%%%%%%%%%%%%%%%%%%
\begin{table}
\begin{center}

\caption{Overall optical depths for the best-fitting models plotted in Fig.\
1. The last entry lists the flux-averaged optical depth $\tau_{\rm F}$. The
other properties of the models are: dust composition, 95 per cent amorphous
carbon and 5 per cent SiC (by mass);  grain size distribution given in
equation (2), with $a_0$ = 0.2 \mic; dust condensation temperature $T_1$ = 750
K; stellar temperature $T_\ast$ = 2200 K.}

\smallskip
\begin{tabular}{@{}cccc}
\hline
 $\lambda$ (\mic) & $\varpi$\n(a) &
 $\tau^{\rm max}$\n(b) & $\tau^{\rm min}$\n(c)       \\
\noalign{\smallskip}
   0.55 &    0.52\0           &  20    &  24      \\
   1.0  &    0.44\0           &  14    &  17      \\
   2.2  &    0.40\0           &   4.7  &   5.7    \\
   5.0  &    0.13\0           &   1.0  &   1.2    \\
   10   &    0.026            &   0.32 &   0.40   \\
  100   & $1.0\times10^{-4}$  &   0.01 &   0.012  \\
   ---  &    ---              &   2.0  &   2.4    \\ \hline
\end{tabular}
\end{center}
\smallskip

(a) Albedo  \\
(b) Total optical depth at maximum light\\
(c) Total optical depth at minimum light\\

\end{table}
%%%%%%%%%%%%%%%%%%%%%%%%%%%%%%%%%%%%%%%%%%%%%%%%%%%%%%%%%%%%%%%%%%%%%%%%%%%%

The parameters of our best-fitting model are summarized in Table 1. Note again
that the fit to the spectral shape $f_\lambda$ is obtained without specifying
the absolute size of the envelope, mass-loss rate, luminosity or distance to
the star.  For given dust grains, the resulting $f_\lambda$ is primarily
determined by the overall optical depth.  Furthermore, the flux scale never
entered the fitting procedure. Actual fluxes are obtained from $f_\lambda$
through simple multiplication by the bolometric flux $F$. Comparison of fluxes
from our best-fitting model with observations gives $F = 2.1\times10^{-8}$ W
m$^{-2}$ at maximum light, in agreement with Sopka et al.\ (1985).  In IE96 we
show that the bolometric flux and angular diameter of the dust condensation
zone, $\theta_1 = 2r_1/D$ where $D$ is the distance to the source, are related
via
\eq{\label{theta}
     \theta_1 = 0.17 \, \alpha
      \left(F \over 10^{-8} {\rm\,W\,m^{-2}}\right)^{\!0.5}
      \left({ 10^3 {\rm \,K}}\over T_1 \right)^{\!2} \hbox{ arcsec},
}
where $\alpha$ is a dimensionless coefficient of order unity characteristic of
the model.  This coefficient, determined theoretically from the overall
solution, depends primarily on the grain optical properties and only slightly
on $T_\ast$, $T_1$ and overall optical depth. From our best-fitting model for
IRC+10216 we find that $\alpha = 1.3$ for this source, and therefore at maximum
light $\theta_1^{\rm max}$ = 0.56 arcsec. With an expected bolometric amplitude
of 1 mag, $\theta_1^{\rm min}$ = 0.35 arcsec at minimum light. These estimates
for the angular scale must agree with high-resolution observations.

\section             {SPATIALLY RESOLVED OBSERVATIONS}

%%%%%%%%%%%%%%%%%%% Figure 2 %%%%%%%%%%%%%%%%%%%%%%%%%%%%%
\begin{figure}
%\centering \leavevmode \figsize \epsfbox[85 75 520 720]{ircvisi.ps}

\caption{Visibility functions for IRC+10216. Lines represent model results,
symbols the observations. Solid lines and full symbols (including + and
$\ast$) correspond to phases close to maximum light, open symbols and dashed
lines to phases close to minimum. Data are from Sutton, Betz \& Storey (1979)
($\star$), Selby, Wade \& Sanchez Magro (1979) ($\diamondsuit$), McCarthy,
Howell \& Low (1980) ($\triangle$), Mariotti et al.\ (1983) ($\circ$), Dyck et
al.\ (1984) (+), Dyck et al.\ (1987) ($\ast$), Benson, Turner \& Dyck (1989)
($\oplus$) and Danchi et al.\ (1990), (1994) ($\Box$). Phases and angular
sizes of the dust condensation zone are listed in Table 2.}

\end{figure}
%%%%%%%%%%%%%%%%%%%%%%%%%%%%%%%%%%%%%%%%%%%%%%%%%%%%%%%%%%%%%%%%%%%%%%%%%%%%

In interpreting the spatially resolved observations of IRC+10216, the source
variability must be taken into account.  As the luminosity varies during the
stellar cycle, the envelope temperature varies too. Therefore, as noted
already by Danchi et al.\ (1990, 1994), the dust condensation radius varies
during the stellar cycle and $r_1$ scales as $L^{0.5}$ (IE96). Because of the
movement of the shell's inner boundary, the overall optical depth is expected
to vary too, so that maximum light has minimum $\tau$. With optical depths
determined from the spectral shape we fit our models to visibility
observations at minimum and maximum light obtained at 2.2, 5 and 10 \mic.  In
these fits, $\theta_1$ is taken as a free parameter, providing an independent
estimate for it.  Fig.\ 2 shows comparison of model results with observations.
Since the data at 2.2 \mic\ are spatially asymmetric, the plotted results are
spatially averaged. Phases and values of $\theta_1$ are summarized in Table 2.
The independent fits for $\theta_1$ from the visibility and the SED agree
within 15--20 per cent on average. It can be estimated that, within 20 per
cent, the angular size of the dust condensation point varies between 0.3 and
0.5 arcsec.

%%%%%%%%%%%%%%%%%%%%%%%%%%%%% Table 2 %%%%%%%%%%%%%%%%%%%%%%%%%%%%%
\begin{table}
\begin{center}

\caption {Angular size of the dust condensation zone at maximum and minimum
light determined from the available visibility observations. References for
the observations are listed in the caption of Fig.\ 2.}

\smallskip
\begin{tabular}{@{}cccc}
\hline
$\lambda$ &  $\Phi$\n(a) & $\theta_1$\n(b) & $\Delta$\n(c) \\
   (\mic) &              &    (arcsec)     &  (per cent)   \\
\noalign{\smallskip}
   2.2    &    max       &       0.40      &      29       \\
   2.2    &    min       &       0.29      &      17       \\
   5      &    min       &       0.32      &     \08       \\
   10     &    max       &       0.45      &      20       \\
   10     &    min       &       0.30      &      14       \\ \hline
\end{tabular}
\end{center}
\smallskip

(a) Phase of the light curve at which visibility
    observations were made at the wavelength listed in the first column \\
(b) Angular size of the dust condensation zone determined
    from model fit to the visibility observations  \\
(c) The percentage difference between the angles listed in column (b) and
those determined from equation (3) and the best fit to the SED
(0.56 arcsec for max and 0.35 arcsec for min).  \\

\end{table}
%%%%%%%%%%%%%%%%%%%%%%%%%%%%%%%%%%%%%%%%%%%%%%%%%%%%%%%%%%%%%%%%%%%%%%%%%%%%%%

Previous models have never achieved simultaneous agreement for both the SED
and spatially resolved observations at short wavelengths (e.g. Martin \&
Rogers 1987).  In all these models, optical depths that fitted the SED
produced a 2.2-\mic\ visibility too large at $q \ga$ 1 arcsec$^{-1}$.  This
problem is directly related to the flux shortage of the models at these
wavelengths and is another manifestation of the need for large grains.  We
have shown in IE96 that the value of the visibility when it levels off at large
$q$ is simply exp($-\tau_{\rm sca}$) for $\lambda \la$ 3 \mic, where
$\tau_{\rm sca}$ is the scattering optical depth. Therefore, $\tau_{\rm sca}$
must be increased to reduce the visibility.  Also, because $\tau_{\rm abs}$ is
fixed from the spectral shape, this increase translates to an increased
albedo, implying the presence of grains with sizes of \about\ 0.2 \mic.  This
independent estimate of the grain sizes provides further support for the one
obtained from the spectral shape.

Recently, Danchi et al.\ (1994) obtained visibility curves for IRC+10216 at 11
\mic\ close to maximum and minimum light.  From these observations they find
$\theta_1 \approx$ 0.1--0.2 arcsec, $\tau_{11}=1.24$ and $T_1=1360$ K. These
results, determined by fitting visibility curves at the single wavelength 11
\mic, differ from ours; by comparison, from the spectral shape we find
$\tau_{11} = 0.3-0.4$ and $T_1 = 750$ K.  It is important to note that the
Danchi et al.\ data, displayed as solid and open squares in the bottom panel
of Fig.\ 2, are properly fitted by our models together with all other data.
By contrast, the Danchi et al.\ fits rely on a limited data set, confined to a
single wavelength and visibilities $<$ 0.4, which does not sufficiently
constrain the model parameters (IE96). Indeed, the values of $\tau_{11}$ and
$T_1$ deduced by Danchi et al.\ cannot produce a simultaneous fit to the
visibilities and SED at wavelengths shorter than 5--6 \mic.

%%%%%%%%%%%%%%%%%%% Figure 3 %%%%%%%%%%%%%%%%%%%%%%%%%%%%%
\begin{figure}
%\centering \leavevmode \figsize \epsfbox[95 120 500 735]{ircblmhf.ps}

\caption{Single-scan (E--W) imaging of IRC+10216 at 10 \mic. The thick solid
line in the top panel is the observations of Bloemhof et al.\ (1988).
Superimposed on it is our model result drawn as a dashed line, hardly
distinguishable from the observations.  It is obtained by a two-dimensional
convolution of the surface brightness for $\theta_1$ = 0.35 arcsec (innermost
thin solid line) with the point-spread function (PSF, dot-dashed line; all
profiles are normalized to unity at the peak).  In the bottom panel, the
dashed line is the surface brightness deduced by Bloemhof et al.\ by
one-dimensional deconvolution of the observed profile with the PSF.  The
dot-dashed line is the one-dimensional convolution of our model result with
the PSF. The thick solid line is the two-dimensional convolution of our model
result for $\theta_1$ = 0.35 arcsec with the PSF (the same as the dashed line
in the top panel). The two thin solid lines below and above this curve
correspond to curves derived analogously from models with $\theta_1$ = 0.30
and 0.40 arcsec, respectively.}

\end{figure}
%%%%%%%%%%%%%%%%%%%%%%%%%%%%%%%%%%%%%%%%%%%%%%%%%%%%%%%%%%%%%%%%%%%%%%%%%%%%

Bloemhof et al.\ (1988) obtained a single-scan image of IRC+10216 at 10 \mic\
close to minimum light (phase $\simeq$ 0.4).  We computed the profile expected
in those observations from the model surface brightness determined for this
phase from the spectral shape. The top panel in Fig.\ 3 shows the comparison
between the observed profile (outermost thick solid line) and our model result
(dashed line, overlapping the observations; all displayed profiles are
normalized to unity at their peaks).\footnote{To remove a slight asymmetry in
the observations, all profiles are symmetrized east--west.} The innermost thin
solid line is the surface brightness distribution obtained from our model with
$\theta_1$ = 0.35 arcsec.  The central peak corresponds to the stellar
contribution and the features at relative RA $\pm \theta_1/2$ to the dust
formation zone. The model result is obtained from a two-dimensional
convolution of this profile with the point-spread function (PSF), shown with
the dot-dashed line. This fit to the observed image provides independent
determination of $\theta_1$ around minimum light, in agreement with the
previous two.

Our model surface brightness is considerably different from the one deduced by
Bloemhof et al., displayed with the dashed thick line in the bottom panel. In
addition to the contribution of a central component with a width of \about\
0.4 arcsec, this profile also requires a more extended component with a width
of \about\ 2 arcsec.  Bloemhof et al.\ obtained this distribution from a one-
rather than a two-dimensional deconvolution of their observed profile with the
PSF.  We have verified that the results of the commonly used one-dimensional
convolution with the PSF are indistinguishable from those of the proper
two-dimensional convolution for centrally peaked surface brightness.  However,
in the case of extended structures whose peaks do not coincide with the centre
of symmetry (e.g., a ring), the two procedures produce different results. A
one-dimensional convolution of our model surface brightness with the PSF
produces the thin dot-dashed line in the bottom panel, considerably different
from the result of the proper two-dimensional convolution, repeated in the
bottom panel as the thick solid line.

The possible existence of an extended \about\ 2 arcsec component was first
conjectured from lunar occultation observations by Toombs et al.\ (1972)
because they could not properly fit their results with a single, sharp-edged
disc with a diameter of \about\ 0.4 arcsec.  Thus they invoked an additional,
larger sharp-edged disc. Recently, Sloan \& Egan (1995) also modelled their
10-\mic\ observations, obtained with $0.9\times2.0$ arcsec$^2$ resolution, in
terms of single-temperature components, a procedure that produced two dust
shells.  One shell ranges in diameter from 0.055 to 0.67 arcsec with a single
temperature of 340 K, the other from 1.5 all the way to 5.2 arcsec with a
single temperature of 240 K.  However, our model, which has no sharp edges or
discrete temperature components, properly explains both the Toombs et al.\ and
Sloan \& Egan observations because it is extended. Therefore, our modelling
does not support the existence of the conjectured 2 arcsec component or any
other discrete shell.  We find no need to augment the steady-state outflow
with any additional components.

Direct imaging is a most sensitive method for determining the dust
condensation radius of optically thin envelopes.  In the bottom panel of Fig.\
3 we display in thin solid lines the imaging results expected with the PSF of
Bloemhof et al.\ when $\theta_1$ is varied by only $\pm$ 0.05 arcsec. Since
the expected variation of $\theta_1$ between minimum and maximum light is
considerably larger, measurements of this variation during the stellar cycle
can provide an important check of our models.

\section                   {DYNAMICAL PROPERTIES}

The quantities determined from IR observations can be used to constrain
dynamical properties of the IRC+10216 outflow. Momentum flux conservation
relates the mass-loss rate $\Mdot = \Mdot_{-5} \times 10^{-5}$ \Mo\,yr$^{-1}$
and $v_{\rm e}$, the terminal outflow velocity in \kms, to radiative properties
via
\eq{
             \Mdot_{-5} v_{\rm e} = 20 \tau_{\rm F} L_4  (1 - \Gamma^{-1}).
}
This relation is valid in steady-state for the time-averages of $\tau_{\rm F}$
and the stellar luminosity $L_\ast = L_4\times10^4~\Lo$, where $\Gamma\
(\propto M_\ast/L_4)$ is the gravitational correction (IE95).  Our model
calculations give $\Gamma^{-1} = 0.13$ for IRC+10216 if the stellar mass is
$M_\ast = 1~\Mo$ and $L_4 = 1.5$; uncertainties in $M_\ast$ and $L_4$ are of
minor importance since $\Gamma^{-1}$ is so small. From our models we find that
$\tau_{\rm F}$ varies from 2 at maximum light to 2.4 at minimum (see Table 1).
With a terminal outflow velocity of 15 \kms\ (Morris, Lucas \& Omont 1985;
Zuckerman \& Dyck 1986), this gives
 \eq{
                             \Mdot_{-5} = 2.1 L_4\,,
}
where $L_4$ refers to the luminosity at maximum light, assuming a bolometric
amplitude of 1 mag.

The dust mass-loss rate $\Mdot^{\rm d}$ is directly related to the optical
depth via
 \eqarray{
 \tau &= &{3 \Mdot^{\rm d} \over 16 \upi v_1 \rho_{\rm s} r_1 } \\
       &&\ \times\int^{\infty}_1\!\! \nhat_{\rm d}(y)\,{\rm d}y
         \int^{\infty}_0{Q(a)\over a}\ n_{\rm J}(a)\,{\rm d}a. \nonumber
 }
Here $\rho_{\rm s}$ is the dust solid density (1.85 g cm$^{-3}$, Rouleau \&
Martin 1991); $Q(a)$ is the extinction efficiency of grains with radius $a$;
$\nhat_{\rm d}$ is the dust density profile normalized to unity at $r_1$; and
$v_1$ is the outflow velocity at $r_1$, assumed to be 1 \kms. This velocity
corresponds to the velocity at the sonic point (Deguchi 1980) and introduces
the principal uncertainty in determining $\Mdot^{\rm d}$. From this result and
expressions for $r_1$ listed in IE96, our model calculations of IRC+10216
produce
 \eq{
               \Mdot^{d}_{-8} = 4.2 \sqrt{L_4}\,,
 }
where $\Mdot^{d}_{-8} = \Mdot^{d}/10^{-8}$ \Mo\ yr$^{-1}$; the variation with
$\sqrt{L_4}$ reflects the dependence of $r_1$ on luminosity (cf.\ Section 3).
The ratio of the mass-loss rates of the entire envelope and the dust component
produces the average gas-to-dust ratio
\eq{
   r_{\rm gd} = 1000 {\Mdot_{-5} \over \Mdot^{\rm d}_{-8}} = 500 \sqrt{L_4}\,.
}

In all these relations, the luminosity remains unknown.  Since the bolometric
flux is $2.1\times10^{-8}$ W m$^{-2}$, the luminosity obeys
 \eq{
                           L_4 = 1.5\, D_{150}^2,
}
where $D = D_{150}\times150$ pc is the distance to the star.  In terms of this
parametrization, the various quantities listed above are
 \eq{
                      \Mdot_{-5} = 3.1\,D_{150}^2\,, \quad
                      \Mdot^{\rm d}_{-8} = 5.2\,D_{150}\,, \quad
                      r_{\rm gd} = 610\,D_{150}\,.
}
These results are in good agreement with independent estimates if we take $D$
= 150 pc. Estimates for $\Mdot_{-5}$ based on CO data range from 1--2 (Jura
1994, and references therein) to 4.7 (Kwan \& Webster 1993).  Values derived
for $\Mdot^{\rm d}_{-8}$ vary from 5.6 (Griffin 1990) to 9.8 (Kastner 1992).
Gas-to-dust ratios for carbonaceous winds in late-type stars range from 260
(Volk, Kwok \& Langill 1992) to 670 (Knapp 1985).

The most-often quoted distance to IRC+10216 is 290 pc, derived by Herbig \&
Zappala (1970) for an assumed luminosity of $L_4 = 5.5$. This distance leads
to unrealistically high mass-loss rates and gas-to-dust ratios. A closer
distance of 100--150 pc has been proposed by Zuckerman, Dyck \& Claussen
(1986) and Kastner (1992), and our results support these suggestions. Based on
theoretical considerations, Martin \& Rogers (1987) pointed out that $L_4 \ge
1.2$. Thus, IRC+10216 is probably not closer than 130 pc, and the value of 150
pc adopted by Jura (1994) and in this work is likely to be close to the true
distance. With this distance of 150 pc, the radius of the dust condensation
zone varies from $\sim 3\times10^{14}$ cm at minimum light to $\sim
5\times10^{14}$ cm at maximum.

\section                         {DISCUSSION}

Our model for IRC+10216 provides a description of the IR observations based on
a self-consistent treatment of the dynamics and radiative transfer.  The model
provides simultaneous agreement for both the SED and all high-resolution
observations of this star.  Particularly encouraging is the agreement between
the three independent determinations of the angular size of the dust
condensation point (\about\ 0.3 arcsec at minimum light). Our results
demonstrate the advantage of following the two-step modelling procedure
outlined in the Introduction.  The most appropriate approach is first to
constrain the overall optical depth by the spectral shape, then use visibility
data to determine the size of the envelope.

Because of the close agreement obtained for such a variety of independent
observations, we do not expect major changes in the parameters determined
here.  The model could still be improved by considering the asymmetry of the
envelope.  As mentioned in Section 2, although symmetric at wavelengths longer
than \about\ 3--4 \mic, the observed surface brightness is slightly asymmetric
at short wavelengths.  Such a dual appearance can be understood in terms of
the basic physical processes that control the IR radiation at the different
spectral regions.  At short wavelengths, scattering dominates the observed
radiation. Since scattering can be expected to map the entire envelope,
scattered radiation should reflect the density distribution, displaying any
asymmetry in it.  As long as the elongation is not severe, it is reflected
only in the shape of the image, not in the flux. On the other hand, radiation
at longer wavelengths is dominated by dust emission, predominantly controlled
by the dust temperature distribution.  For slightly elongated density
distributions, the dust temperature distribution can still be spherically
symmetric to a good degree of approximation because it is mostly controlled by
the distance from the central star.  This explains the close agreement of our
spherically symmetric model with the data. A slightly enhanced mass-loss rate
in the equatorial plane, as proposed by Ridgway \& Keady (1988) and supported
by Kastner \& Weintraub (1994), can be accommodated without a significant
effect on our results. On the other hand, our model cannot describe the
emission from a bipolar nebula, the geometry suggested for IRC+10216 by Dyck
et al.\ (1987), and its success indicates that such a drastic departure from
spherical symmetry may not be necessary to explain the observations.

We resolve the difficulties encountered in previous studies at short
wavelengths by including large grains.  The short-wavelength behaviour of both
the SED and the visibility shows that grains as large as \about\ 0.15--0.2
\mic\ are present. Jura (1994) recently discussed the grain size distribution
for IRC+10216, based on polarization in the {\it K} band and shielding of
circumstellar molecules against destruction by interstellar UV radiation. He
finds that grains as large as \about\ 0.1 \mic\ exist in the outer envelope
(more than 15 arcsec from the star), in good agreement with the sizes obtained
here. Since our analysis of grain sizes is primarily affected by the inner
regions of the envelope while Jura's results apply to the outer regions, we
conclude that the grain sizes do not change significantly through the
envelope. The effects of sputtering, grain growth, etc., do not seem to be too
important after the initial dust formation.

\section*{Acknowledgments}

We thank E.\ Bloemhof and R.\ Danen for providing us with observations of
IRC+10216.  We are especially indebted to E.\ Bloemhof for his generous help
in clarifying the analysis of the data. We also thank the referee, C.\
Skinner, for his careful reading and useful comments.  This research has made
use of the {\sc SIMBAD} database, operated at CDS, Strasbourg, France, and the
{\sc ADS} database. Support by NSF grant AST-9321847, NASA grant NAG 5-3010
and the Center for Computational Sciences of the University of Kentucky is
gratefully acknowledged.

\label{lastpage}

\end{document}